\documentclass{ws-ijmpd}

\def\eqiaz{\begin{eqnarray*}}
\def\eqfaz{\end{eqnarray*}}
\def\eqia{\begin{eqnarray}}
\def\eqfa{\end{eqnarray}}
\def\btab{\begin{tabular}}
\def\etab{\end{tabular}}
\def\bar{\begin{array}}
\def\ear{\end{array}}
\def\rfr#1{Equation (\ref{#1})}

\def\Rfr#1{Equation (\ref{#1})}

\def\eqi{\begin{equation}}
\def\eqf{\end{equation}}

\def\lb#1{\label{#1}}

\begin{document}

\def\nocropmarks{\vskip5pt\phantom{cropmarks}}

\let\trimmarks\nocropmarks

\markboth{Lorenzo Iorio}
{A gravitomagnetic effect}

\catchline{}{}{}

\title{A GRAVITOMAGNETIC EFFECT ON THE ORBIT\\ OF A TEST BODY
DUE TO\\ THE EARTH'S VARIABLE ANGULAR MOMENTUM}

\author{\footnotesize LORENZO IORIO}

\address{Dipartimento di Fisica dell'Universit$\grave{\rm a}$ di Bari\\
Via Amendola 173, 70126, Bari, Italy}

\maketitle

\begin{abstract}
The well known general relativistic Lense-Thirring drag of the
orbit of a test particle in the stationary field of a central
slowly rotating body is generated, in the weak-field and
slow-motion approximation of General Relativity, by a
gravitomagnetic Lorentz-like acceleration in the equations of
motion of the test particle. In it the gravitomagnetic field is
due to the central body's angular momentum supposed to be
constant. In the context of the gravitational analogue of the
Larmor theorem, such acceleration looks like a Coriolis inertial
term in an accelerated frame. In this paper the effect of the
variation in time of the central body's angular momentum on the
orbit of a test mass is considered. It can be shown that it is
analogue to the inertial acceleration due to the time derivative
of the angular velocity vector of an accelerated frame. The
possibility of detecting such effect in the gravitational field of
the Earth with LAGEOS-like satellites is investigated. It turns
out that the orbital effects are far too small to be measured.
\end{abstract}

\section{Introduction} In the weak-field and slow-motion
approximation of General Relativity, a test particle orbiting a
central slowly rotating body of mass $M$ and angular momentum
${\bf J}$, assumed to be constant, is acted upon by a non-central
acceleration$^1$ of the form \eqi{\bf a}_{\rm
GM}=\frac{{\bf v}}{c}\times {\bf B}_{\rm g}\lb{eqb},\eqf in which
${\bf v}$ is the velocity of the test particle, $c$ is the speed
of light in vacuum and ${\bf B}_{\rm g}$ is the gravitomagnetic
field given by$^2$ \eqi{\bf B}_{\rm
g}=\frac{2G}{c}\frac{\left[{\bf J}-3\left({\bf J}\cdot
\hat{r}\right)\hat{r}\right]}{r^3}\lb{gmf}.\eqf In it $\hat{r}$ is
the unit position vector of the test particle and $G$ is the
Newtonian gravitational constant. The gravitomagnetic field ${\bf
B}_{\rm g}$ can be derived$^2$ from the gravitomagnetic potential
${\bf A}_{\rm g}$ \eqi{\bf A}_{\rm
g}=\frac{G}{c}\frac{{\bf J}\times{\bf r}}{r^3}\lb{acca}\eqf
according to \eqi\nabla\times {\bf A}_{\rm g}=-\frac{{\bf B}_{\rm
g}}{2}.\eqf  \Rfr{gmf} and \rfr{acca} hold far from the central
body supposed to be spherically symmetric and rigidly rotating. An
inertial frame $K\{x,y,z\}$ with its origin located at the center
of mass of the rotating body, the $z$ axis directed along ${\bf
J}$ and the $\{x,y\}$ plane equal to the equatorial plane of the
gravitating source is adopted.

This effect, which affects the longitude of the ascending node $\Omega$ and the
argument of pericenter $\omega$
of the orbit of a test particle$^3$, is currently under measurement by
analyzing the laser-ranged data to LAGEOS
satellites in the gravitational field of the Earth$^4$.

According to the gravitational analogue of the Larmor theorem$^5$,
we could obtain \rfr{eqb} by considering an
accelerated frame rotating with angular velocity
\eqi\vec{\Omega}_{\rm L}= \frac{{\bf B}_{\rm g}}{2c}.\eqf Indeed,
in it an inertial Coriolis acceleration \eqi{\bf a}_{\rm
Cor}=2{\bf v}\times\vec{\Omega}_{\rm L}\lb{corio}\eqf is
experienced by the proof mass.\footnote{It should be noticed that
the equivalence is useful only in spatial regions in which the
field considered is uniform, so that$^5$ $-4{\bf A}_{\rm g}={\bf
B}_{\rm g}\times{\bf r}$.}

It seems natural to pose the following question. In an accelerated
frame, apart from the Coriolis and centrifugal inertial
forces\footnote{The centrifugal force is proportional to the
square of $\Omega_{\rm L}$, so that the impact of a possible
gravitomagnetic analogue, proportional to the square of ${\bf
B}_{\rm g}$, can be neglected because we are in the weak-field
approximation.}, a particle feels an acceleration \eqi {\bf
r}\times\frac{{\rm d}\vec{\Omega}_{\rm L}}{{\rm d} t}\eqf as well
if the angular velocity vector $\vec{\Omega}_{\rm L}$ is
time-dependent. Does a gravitomagnetic analogue of such term
exist? If so, it should be induced by temporal variations of the
angular momentum of the central body ${\bf J}$. In the case of the
Earth we know that ${\bf J}_{\oplus}$ changes in time due to the
luni-solar torques which generate, among other things, the
precession of equinoxes$^6$.

In this paper we will try to investigate this feature. The plan of
the work is as follows. In Section 2 we derive the full expression
of the gravitomagnetic acceleration experienced by the test mass
when the gravitomagnetic potential due to the central rotating
mass is time-dependent. In Section 3 the orbital effects on the
semimajor axis, the inclination and the node of the orbit of an
Earth satellite are derived and the possibility of measuring such
effect in the terrestrial field with LAGEOS-like satellites are
discussed. Section 4 is devoted to the conclusions.

In Table 1 the numerical values of the parameters used in the text
are quoted. For the Earth, the Moon and the Sun reference$^7$ has
been used.

\begin{table}[htbp]
\ttbl{30pc}{Parameters used in the text.}
{\begin{tabular}{lccc}\\
\multicolumn{4}{c}{}\\[6pt]\hline
Parameter  & Description & Value & Units \\ \hline
$M_{\odot}$ & Mass of the Sun & $1.99\times 10^{33}$ & g \\
$d_{\odot}$ & Earth-Sun mean distance & $1.496\times 10^{13}$ & cm \\
$M_{\rm m}$ & Mass of the Moon & $7.3483\times 10^{25}$ & g \\
$d_{\rm m}$ & Earth-Moon mean distance & $3.844\times 10^{10}$ & cm \\
$(C-A)_{\oplus}$ & Earth's principal moments of inertia &
$2.6249\times 10^{42}$
& g cm$^2$ \\
$G$ & Newtonian constant of gravitation & $6.673\times 10^{-8}$ &
g$^{-1}$ cm$^3$ s$^{-2}$\\
$c$ & Speed of light in vacuum & $2.9979\times 10^{10}$ & cm
s$^{-1}$ \\
$\varepsilon$ & Obliquity of ecliptic & 23.44 & deg \\
$a_{\rm LAGEOS}$ & LAGEOS semimajor axis & $1.2270\times 10^9$ &
cm\\
$e_{\rm LAGEOS}$ & LAGEOS eccentricity & 0.0045 & -\\
$i_{\rm LAGEOS}$ & LAGEOS inclination & 110 & deg\\
$n_{\rm LAGEOS}$ & LAGEOS mean motion & $4.643\times 10^{-4}$ & rad
s$^{-1}$\\
\hline
\end{tabular}}
\end{table}

\section{The gravitomagnetic acceleration}

In the weak-field and slow-motion linearized approximation of
General Relativity, the Lagrangian ${\mathcal{L}}$ of a test
particle of mass $m$ moving in a metric
$ds^2=g_{00}({dx^0})^2+g_{ii}({dx^i})^2 +2g_{0k}dx^0 dx^k$ is given by$^2$
\eqi{\mathcal{L}}={\mathcal L}_{({\rm
GE})}-\frac{2m}{c}{\bf A}_{\rm g}\cdot{\bf v},\lb{lagr}\eqf where
the gravitomagnetic potential ${\bf A}_{\rm g}$ is generated by
the off-diagonal components $g_{0k}$, $k=1,2,3$ of the space-time
metric while ${\mathcal L }_{({\rm GE})}$ denotes the
gravitoelectric Schwarzschild terms. The Lagrange equations of
motion are \eqi\frac{{\rm d}}{{\rm d}
t}\left(\frac{\partial{{\mathcal{L}}} } {\partial{\bf
v}}\right)-\frac{\partial{{\mathcal{L}}}}{\partial{\bf r}}=0.\eqf
From \rfr{lagr} it follows
\eqia\frac{{\rm d}}{{\rm d} t}\left(\frac{\partial{{\mathcal{L}}}
} {\partial{\bf v}}\right) & = & m\frac{{\rm d}{\bf v}}{{\rm d}
t}-\frac{2m}{c}\left({\bf v}\cdot\nabla\right){\bf A}_{\rm g}
-\frac{2m}{c}\frac{\partial{\bf A}_{\rm g}}{\partial t}+(\rm GE),\\
\frac{\partial{{\mathcal{L}}}}{\partial{\bf r}} & = &
-\frac{2m}{c}\nabla\left({\bf A}_{\rm g}\cdot{\bf v}\right)+(\rm
GE).\eqfa Then, the equations of motion of a test particle are
\eqia m\frac{{\rm d}{\bf v}}{{\rm d} t} &=&
-\frac{2m}{c}\nabla\left({\bf A}_{\rm g}\cdot{\bf v}\right)
+\frac{2m}{c}\left({\bf v}\cdot\nabla\right){\bf A}_{\rm g}
+\frac{2m}{c}\frac{\partial{\bf A}_{\rm g}}{\partial t} + (\rm GE)=\nonumber\\
&-& \frac{2m}{c}\left[{\bf v}\times\left(\nabla\times{\bf A}_{\rm
g}\right)\right] +\frac{2m}{c}\frac{\partial{\bf A}_{\rm
g}}{\partial t}+ (\rm GE).\lb{equ} \eqfa In \rfr{equ} $(\rm GE)$
represents the post-Newtonian gravitoelectric acceleration due to
the Schwarzschild part of the metric, of order ${\mathcal
O}(c^{-2})+{\mathcal O}(c^{0})$, which reduces to the Newtonian
central term, of order ${\mathcal O}(c^{0})$, in the limit
$c\rightarrow\infty$. For $-4{\bf A}_{\rm g}={\bf B}_{\rm
g}\times{\bf r}$, i.e. when the field is uniform, the
gravitomagnetic part of \rfr{equ} becomes \eqi{\bf a}_{\rm
GM}=\frac{{\bf v}}{c}\times{\bf B}_{\rm
g}-\frac{1}{2c}\frac{\partial{\bf B}_{\rm g}}{\partial
t}\times{\bf r}.\lb{dbg}\eqf If we use \rfr{corio}, \rfr{dbg}
becomes \eqi {\bf a}_{\rm GM}=2{\bf v}\times\vec{\Omega}_{\rm
L}+{\bf r}\times \frac{\partial\vec{\Omega}_{\rm L}}{\partial
t}.\lb{lar} \eqf \Rfr{lar} shows that the gravitational analogue
of the Larmor theorem extends exactly also to the case in which
the gravitomagnetic field is explicitly time-dependent.

\section{The consequences of the variability of Earth's angular
momentum}

We will focus on the second term of \rfr{dbg}. When the
mass-energy distribution is a central spherically symmetric
rigidly rotating body, so that \rfr{gmf} and \rfr{acca} can be
applied, \rfr{dbg} can be written \eqi {\bf a}_{\rm
Lar}=\frac{2G}{c^2 r^3}\frac{{\rm d}{\bf J}}{{\rm d} t}\times{\bf
r}= \frac{2G}{c^2 r^3}{\bf M}\times{\bf r},\lb{mom}\eqf in which
${\bf M}=\frac{{\rm d}{\bf J}}{{\rm d} t}$ is the momentum of the
external forces. Indeed, in a body--fixed rotating frame with an
angular velocity $\vec{\omega}_0$ the dynamical Euler equations
hold \eqi\frac{{\rm d}{\bf J}}{{\rm d}t}=\frac{\partial{\bf
J}}{\partial t}+\vec{\omega}_0\times{\bf J}={\bf M}.\eqf However,
we are using an inertial frame so that \eqi\frac{{\rm d}{\bf
J}}{{\rm d}t}=\frac{\partial{\bf J}}{\partial t}={\bf M}.\eqf In
the case of the Earth ${\bf M}$ is the external torque exerted on
its equatorial centrifugal bulge by the other bodies of the Solar
System$^8$. We will consider only the Moon and the
Sun\footnote{Recall that the other planets induce secular effects
on the ecliptical plane: the planetary precession and a secular
change in the obliquity $\varepsilon$. The luni-solar torque
affects the equatorial plane causing the Earth's spin to precess
and nutate about the ecliptic pole. While the precession does not
affect $\varepsilon$ but only the position of the equinox in the
ecliptic, the nutation does affect $\varepsilon$ periodically.}.
In the frame $K$ previously defined and by assuming the Earth as
an oblate spheroid whose rotation is affected by the presence of a
disturbing body of mass $M_{\rm B}$ (e.g. the Sun) moving around
it on an approximately circular orbit of radius $d$ lying on the
ecliptic plane, the external torque can be written as$^8$
\eqia M_x & = & \frac{3GM_{\rm
B}(C-A)_{\oplus}}{d^3}\sin^2\lambda\sin\varepsilon\cos\varepsilon\lb{m1}\\
M_y & = & -\frac{3GM_{\rm
B}(C-A)_{\oplus}}{d^3}\sin\lambda\cos\lambda\sin\varepsilon\\
M_z & = & 0,\lb{m3} \eqfa where $(C-A)_{\oplus}=J_2
M_{\oplus}R^2_{\oplus}$ in which $J_2$ is the first even zonal
coefficient of the multipolar expansion of the terrestrial
gravitational field and $M_{\oplus}$ and $R_{\oplus}$ are the mass
and the equatorial radius, respectively, of the Earth. Moreover,
$\lambda$ is the ecliptical longitude of the perturbing body and
$\varepsilon$ is the inclination of the ecliptic to the equator.

The next steps in order to evaluate the observable consequences on
the orbit of an Earth artificial satellite consist of projecting
the perturbing acceleration onto the usual radial, along and
cross-track directions, evaluating the so obtained components
$R,T$ and $N$ of ${\bf a}_{\rm Lar}$ in terms of the osculating
Keplerian elements of the orbiter and calculating their rates of
changes by averaging over an orbital revolution of the satellite.
In performing the orbital averages $d$ and $\varepsilon$ will be
considered constant.

From \rfr{mom} and \rfr{m1}--\rfr{m3} it is straightforward to
obtain \eqia
R & = & 0\lb{radial}\\
T & = & 6\left(\frac{G}{c}\right)^2\frac{M_{\rm B}(C-A)_{\oplus}\sin
i}{d^3 r^2}{\mathcal{T}}\lb{along}\\
N & = & -6\left(\frac{G}{c}\right)^2\frac{M_{\rm
B}(C-A)_{\oplus}}{d^3 r^2}\left[
{\mathcal{A}}\cos(\omega+f)+{\mathcal{B}}\sin(\omega+f)\right]\lb{cross},
\eqfa where \eqia {\mathcal{T}} & = & \frac{\sin
2\varepsilon\sin\Omega(1-\cos 2\lambda)}{4}+
\frac{\sin\varepsilon\cos\Omega\sin 2\lambda}{2}\\
{\mathcal{A}} & = & \frac{\sin 2\varepsilon\cos i\sin\Omega(\cos
2\lambda-1)}{4}-
\frac{\sin\varepsilon\cos i\cos\Omega\sin 2\lambda}{2}\\
{\mathcal{B}} & = & \frac{\sin 2\varepsilon\cos\Omega(\cos
2\lambda-1)}{4} +\frac{\cos\varepsilon\sin\Omega\sin 2\lambda}{2},
\eqfa in which $i$ is the inclination of the satellite's orbit and
$f$ is its true anomaly. From \rfr{radial}--\rfr{cross} it can be noticed that the disturbing
acceleration is non-central.

Among the Keplerian orbital elements of a typical Earth satellites the semimajor
axis $a$, the the inclination $i$ and the node $\Omega$
are, in general, well measured, as in the case of LAGEOS laser-ranged
satellites. Then, we will calculate the rates
of change of such elements according to$^6$
\eqia \frac{{\rm d}a}{{\rm d}t}&=&
\frac{2}{n\sqrt{1-e^2}}\left[Re\sin(\omega+f)+T\frac{a(1-e^2)}{r}\right],\lb{sma}\\
\frac{{\rm d}i}{{\rm d}t}&=&
\frac{1}{na\sqrt{1-e^2}}N\frac{r}{a}\cos(\omega+f),\\ \lb{ink}
\frac{{\rm d}\Omega}{{\rm d}t}&=& \frac{1}{na\sqrt{1-e^2}\sin
i}N\frac{r}{a}\sin(\omega+f),\lb{nod} \eqfa in which $e$ is the
satellite's orbital eccentricity, $n=\sqrt{GM a^{-3}}$ is the
satellite mean motion and \eqi r=\frac{a(1-e^2)}{1+e\cos f}\eqf on
the unperturbed Keplerian ellipse.

By inserting \rfr{radial}--\rfr{cross} in \rfr{sma}--\rfr{nod}, averaging over
an orbital revolution of the satellite
and neglecting terms of order ${\mathcal{O}}(e)$, it is possible to obtain the
long-term evolutions of $a$, $i$ and $\Omega$
\eqia \frac{{\rm d}a}{{\rm d}t}&=&
6\left(\frac{G}{c}\right)^2\frac{M_B (C-A)_{\oplus}\sin
i\sin\varepsilon}{na^2 d^3}
\left[\cos\varepsilon\sin\Omega-\right.\nonumber\\
&-&\left.\left(\frac{1+\cos\varepsilon}{2}\right)\sin(\Omega-2\lambda)
+\left(\frac{1-\cos\varepsilon}{2}\right)\sin(\Omega+2\lambda)\right], \lb{dadt} \\
\frac{{\rm d}i}{{\rm d}t}&=&
6\left(\frac{G}{c}\right)^2\frac{M_B (C-A)_{\oplus}\cos
i\sin\varepsilon}{4na^3 d^3} \left[\cos\varepsilon\sin\Omega-\right.\nonumber\\
&-&
\left.\left(\frac{1+\cos\varepsilon}{2}\right)\sin(\Omega-2\lambda)
+\left(\frac{1-\cos\varepsilon}{2}\right)\sin(\Omega+2\lambda)\right],\\
\lb{didt}
\frac{{\rm d}\Omega}{{\rm d}t}&=&
6\left(\frac{G}{c}\right)^2\frac{M_B
(C-A)_{\oplus}\sin\varepsilon}{4na^3 d^3 \sin i}
\left[\cos\varepsilon\cos\Omega+\right.\nonumber\\
&+&\left.\left(\frac{1-\cos\varepsilon}{2}\right)\cos(\Omega-2\lambda)
-\left(\frac{1+\cos\varepsilon}{2}\right)\cos(\Omega+2\lambda)\right].\lb{dodt}\eqfa

From an inspection of \rfr{dadt}--\rfr{dodt} it can be noticed
that there are no secular, linear trends but only long-period
harmonic perturbations whose frequencies are linear combinations
of the longitude of the satellite's ascending node $\Omega$ and
the ecliptical longitude $\lambda$ of the disturbing body $B$.
E.g., for the Sun and LAGEOS we have
\eqia P(\Omega) &=& 1043.67\ {\rm days},\\
P(\Omega-2\lambda) &=& -221.34\ {\rm days},\\
P(\Omega+2\lambda) &=& 155.42\ {\rm days}.
\eqfa
Notice also the $d^{-3}$ dependence, typical of a tidal effect.

Concerning the possibility of measuring such effects in the
terrestrial field, unfortunately there is no hope of detecting
them: indeed, they are $c^{-2}$ effects multiplied by $G^2$. For
LAGEOS the experienced acceleration due to the solar torque is of
the order of \eqi a^{\odot}_{\rm Lar}=
6\left(\frac{G}{c}\right)^2\frac{M_{\odot}(C-A)_{\oplus}}{d_{\odot}^3
a^2}\sim 10^{-17}\ {\rm cm}\ {\rm s}^{-2},\eqf while the present
level of sensitivity is$^9$ $10^{-10}$ cm s$^{-2}$.
Consequently, the amplitudes of the periodic signals of the
semimajor axis, the inclination and the node amounts to $10^{-7}$
cm and $10^{-8}$ mas, respectively\footnote{In integrating
\rfr{dadt}--\rfr{dodt} the corrections $\Delta\lambda$ and
$\Delta\varepsilon$ due to precession and nutation have been
neglected.}. If the action of the lunar torque is considered, it
turns out that \eqi \frac{a^{\rm m}_{\rm Lar}}{a^{\odot}_{\rm
Lar}}\sim 2.1.\eqf A posteriori, this result justifies the
previous choice of neglecting terms of order ${\mathcal{O}}(e)$ in
the satellite's eccentricity and the contribution of the torques
of the other planets of the Solar System.

\section{Conclusions} Motivated by the gravitational analogue of
the Larmor theorem, in this paper we have investigated some
consequences of the variability of the proper angular momentum of
a central slowly rigidly rotating body due to external torques on
the gravitomagnetic equations of motion of a test mass in the
weak-field and slow-motion approximation of General Relativity.
The test particle turns out to be affected by a post-Newtonian
${\mathcal{O}}(c^{-2})$ non-central acceleration analogue to the
inertial acceleration due to the temporal derivative of the
angular velocity vector arising in a non-inertial rotating frame.
In view of a possible measurement of such effect with LAGEOS-like
satellites in the gravitational field of the Earth, we have
calculated the perturbations induced on the semimajor axis, the
inclination and the node of an orbiting proof mass. We have found
no secular terms but only long-term periodic effects: their
magnitude is far too small to be detected.

\section*{Acknowledgements} I wish to thank B. Mashhoon for its
encouragement, the useful material sent to me and his kind
suggestions.

\end{document}